\documentstyle[preprint,aps]{revtex}
\begin{document}
\draft

\title{\bf{Application of the Interface Approach  in  Quantum Ising models}}

\author{Parongama Sen}

\address{Institute of Theoretical Physics,
University of Cologne,}
\address {Zulpicher Strasse 77,
50937  Cologne, Germany }
\address {e-mail paro@thp.uni-koeln.de}
\maketitle
\narrowtext
\begin{abstract}

We investigate  phase transitions
in the Ising model and the ANNNI model in transverse field using the
interface approach. The exact 
result of the Ising chain in  a transverse field 
 is reproduced. 
We find  that apart from the  
   interfacial energy, there are two other 
response functions 
 which show simple scaling behaviour.
For the ANNNI model in a transverse field, the phase diagram  
can be fully studied in the region where a ferromagnetic to
paramagnetic phase transition occurs. The other region  can
be studied partially; the boundary   where the  antiphase vanishes
can be  estimated.

\end{abstract}

\pagebreak
\section  {\bf{Introduction}}

Phase transitions in Ising spin systems driven entirely by 
quantum fluctuations have been getting a lot of attention 
recently [1]. The simplest of such systems is the Ising model
in a transverse field which can be exactly solved in one
dimension. Quantum fluctuations in Ising systems with more 
complicated interactions 
which, for example, incorporate  frustration and or disorder, give rise
to novel and intriguing features.  Recently,  the experimental 
realisation of some cases  like the spin glass system in a 
transverse or tunnelling field, have added to the interest
in such systems [1].

We apply the method of interfaces [2] in the Ising model and the 
anisotropic next
nearest neighbour Ising (ANNNI)  model [3] in a transverse field at zero
temperature to study 
the quantum fluctuation driven transitions. In the process, we also 
explore the
scope  of the so called twist method [2,4]  which we have  shown to have
additional features apart from the ones already known.

Recently, it has been shown in a  variety of spin systems how the 
interfaces caused by twisting a system is closely linked to 
the phase transition. Apart from the application of the twist method to 
several classical models like Ising spins systems, Potts model and 
spin glasses [2], very recently it  has been used for quantum ground
state problems also [4].
  In this method, the interface free energy is generated by the 
excess free energy 
between systems with and without a twist. In general, twisting the
system may be done by changing the boundary condition in 
one direction. The idea is that long range order produces stiffness.
The interface free energy, which is the response to the stress generated
by the twist provides direct information on the stiffness of the ordered
state. For classical systems, 
i.e., in a thermally driven phase transition, 
this method analyzes size ($L$)  and temperature
($T, T_c $ the critical temperature)
 dependence of the stiffness free energy  (which is the increment of
free energy due to the change $\phi $ in boundary conditions) defined by

$$ \Delta F = F_\phi - F_0 ,\eqno{(1)}$$
where $F_\phi$  and $ F_0$ are the free energy with and without twist
respectively.
$\Delta F$ has the simple scaling form [5,2] 

$$ \Delta F = A((T-T_c)L^{1/\nu}) L^{\alpha (T)}\eqno{(2)}$$
where the stiffness exponent $\alpha $ is a  
constant for $T < T_c$,  equal to zero for $T = T_c$  and
  negative for $T > T_c$. Hence   the critical point can be
obtained from $\alpha (T_c) = 0$. In Ising spin  systems with nearest neighbour
interactions, $\alpha   = d -1$ where $d$ is the dimension of the system. 
For frustrated systems, $\alpha$ may be nonintegral [2].

On the other hand, in phase transitions driven by quantum 
fluctuations at zero temperature, one needs to  consider only the
ground state energy (which is equivalent to the free energy) and
here the interfacial free energy 
is expected to have a different stiffness exponent. 

We have applied the twist method in two quantum systems :
first to reproduce  the exact result
of the Ising chain in a transverse field [6] and then to the 
ANNNI  model in a transverse field [1].
In the latter, there are additional frustration effects which
have to be taken under consideration. Our results show that apart from
the interfacial free energy, there are at least two other response 
functions  which carry information of the phase transition and 
follow simple scaling laws. 
In section II, we describe  the method used to study  the quantum Ising models
as well as the results. The results are discussed in section III.

\section  {Method and Results}

The stiffness exponent for the  
quantum model at zero temperature 
is defined in the same way as in (2), the 
role of temperature now being assumed by the transverse field such that

$$ \Delta E = E_0 - E_\phi  $$
$$ = g((\Gamma -\Gamma _c)L^{1/\nu})L^{\phi (\Gamma )}. \eqno{(3)}$$

The Ising chain  in a  transverse field is described by the 
Hamiltonian

$$  H = - J\sum_{i=1}^{L} S_{i}^zS_{i+1}^z
 -\Gamma \sum_{i=1}^{L} S_{i}^{x}\eqno{(4)}$$
and the ferromagnetic to paramagnetic phase transition occurs at $\Gamma /J
= 1$ for $S^z = \pm 1$. 
We take the basis states to be  diagonal in the representation of $S^z$. 
The twist is applied in the following way [2] : in one case we have  fixed
spins pointing parallely  in the left and right boundaries which 
favours the 
ferromagnetic alignment and is called the favourable boundary condition (FBC),
while in the other case we have fixed spins at the boundaries 
antiparallely oriented 
(unfavourable boundary condition or UBC).
The latter generates an interface and hence the excess energy.
 The first spin also interacts with the extra spin (fixed) on its left 
and the last ($L$th) spin interacts with the extra ($L+1$th) spin (fixed)  on 
its right. 

It needs to be clarified here that we have used open
boundary conditions with two extra spins pointed either parallely or 
antiparallely at the edges. This, while generating the interface, 
will  also introduce boundary effects (finite size 
effects in a numerical study):  the two effects are 
intermingled and difficult to separate. It might be possible to
study the interface effect alone by using 
periodic
and antiperiodic boundary conditions [7],  but that involves more
complicated  programming and computer time. 
Therefore,  we have both interface
and boundary  effects, and when we talk of interface effect 
in the rest of the paper, it essentially
includes boundary effect, the latter diminishing  with system size.   

 We proceed to find out the ground state of a system of $L$  spins (excluding 
the two at the boundary) in a
transverse field by using a Lanczos algorithm for both kinds of boundary
conditions (FBC and UBC).

Apart from the interfacial energy defined in 
eq. (3), we also investigate the behaviour of the  interfacial  cooperative
energy  and the interfacial magneitsation. These two quantities 
are defined in the following way :
let $E^c$  = $<\psi|H^c|\psi>$ where $H^c$ is the term(s) in the 
Hamiltonian involving  only the cooperative interaction energy  and 
$|\psi >$  the ground state. For (4), $H^c = -J\sum_{i}^{L} S^{z}_iS^{z}_{i+1}$.
Then the interfacial cooperative energy is given
by 

$$ \Delta E^c =  E{^c}_0 - E^c_\phi  \eqno{(5)}$$.

The interfacial magnetisation is similarly defined 

$$ \Delta m =  m_o - m_\phi  \eqno{(6)}$$
where $m_0 ~(m_\phi )$ is the magnetisation in the ground state with (without)
twist.
We have obtained results for system sizes $L$ = 6 to $L$ = 20 and 
studied the bahaviour of $\Delta E$, $\Delta E^c$ and $\Delta m$. All three
scale in general as (3) giving the exact result $\Gamma _c/J = 1$ and
$\nu = 1 $ (see Fig. 1).
 Although the exact
critical point is known for (4), certain other features are available
from our study which shows novel features of the stiffness 
exponent for quantum systems.  
 We have discussed these
scaling behaviour and commented about them in section III.

We next extend the study to the ANNNI chain in a transverse field. 
The Hamiltonian is described by 

$$  H = - J[\sum_{i=1}^{L} S_{i}^zS_{i+1}^z
 - \kappa \sum_{i=1}^{L} S_{i}^zS_{i+2}^z +\Gamma \sum_{i=1}^{L} S_{i}^{x}]\eqno{(7)}$$
 Here $\kappa $ denotes
the frustration parameter.
The classical ground state without $\Gamma $ at zero temperature 
is exactly known :
ferromagnetic for $\kappa  < 0.5$, antiphase for $\kappa > 0.5$
and highly degenerate phases exist at $\kappa = 0.5$ [3].
The quantum ANNNI model, which is perhaps the simplest model incorporating both
frustration and quantum fluctuation, has been studied extensively 
(and the  corresponding classical model) in the last few years [1].
However, the nature of the ground state and the phase transition is yet
 to be understood clearly especially in the region $\kappa > 0.5$.
It is believed that a floating phase exists [1,8] close to the $\kappa = 0.5$
region which has also been found for the classical two dimensional
model in the free fermion approximation [3]. 
 All earlier studies indicate that there is a ferromagnetic to
paramagnetic transition at $\kappa < 0.5$. Hence, the twist method
is easily applicable here in the same manner as in the 
nearest neighbour Ising case.
In order to impose  favourable and unfavourable boundary conditions,
we fix two spins on the left and right end of the chain,
and find the ground states. 
The spins on the boundaries interact with the extra fixed spins as in the 
Ising case,
with open boundary conditions  prevailing.
For $\kappa < 0.5$, the FBC consists 
of parallel spins, and for UBC, it is antiparallel just like
the nearest neighbour case. It maybe mentioned that one could
do without  bringing in two fixed spins but we keep this in order
it is consistent with the ground states also at $\kappa > 0.5$.
We have applied here the twist method and found that it gives
consistent results in the $\kappa < 0.5$ region where a ferromagnetic
to paramagnetic transition occurs.  Again we find that $\Delta E$,
$\Delta E^c$ and $\Delta m$ have simple scaling forms and we get
the critical field for any $\kappa < 0.5$ in this way. As an example,
we have shown the scaling of the three quantities in Fig. 2 for $\kappa = 0.3$.

In the $\kappa > 0.5$ region, we have no clear idea about what 
kind of a transition is taking place which is clear-cut ferromagnetic to 
   paramagnetic  
in the $\kappa < 0.5$ region. Therefore, all we have attempted to do here
is to find out the phase boundary where the antiphase disappears by
putting appropriate  UBC and FBC for the antiphase. 
 However, there  still 
remains a   problem.  The frustration effects now become dominant and 
the ground state is no longer trivially degenerate. This generates
not a single interface but maybe more than one. Also, because of
the structure of the degenarate ground states due to  the presence of
both nearest and next nearest neighbour interactions, the so called 
unfavourable boundary condition for one particular ground state may become 
favourable for another degenerate ground state, thus making it
difficult to feel the effect of the field due to the twist. 
For 
example, if we set the two spins on the left boundary down and the 
two on the right up, then the state with minimum interaction 
energy is $|uudd....uudd>$,  a member of the set of the 4 degenerate 
ground states in 
the antiphase.  Setting all the boundary spins on the left and right down to 
provide the necessary twist, the new ground  state  should apparently 
have a structure $|uu........uu>$, where we do not know how  
the spins in the interior are oriented.
The cooperative   energy 
contribution at  the boundary to  this state is    
$2J - 4J\kappa$.
However, if we look at another antiphase state 
 which is $|duudd....dduud>$, then the energy contribution at the 
boundary is $- 2J$. Hence it is possible that the latter is lower in 
energy compared to $|uu.......uu>$ especially if 
$-2J  < 2J - 4J\kappa $ or  when $\kappa  < 1. $
Hence, a second antiphase state becomes the ground state 
when the twist is applied therefore making the present method 
ineffective. 
However, with   the quantum term  also present, we observed  from 
the numerical exercise that this  problem disappears  for $\kappa \geq 0.7$ 
where
we find out the phase boundary.
The  interfacial magnetisation is of course not meaningful here. 

 We have estimated the phase boundary where the $<2>$ phase disappears
 again from the  best  scaling  plots for $\Delta E$ and 
$\Delta E_c$ (the $\kappa $ = 1.0 case is shown in Fig. 3). 
However, the data collapse is not so impressive as in the 
$\kappa < 0.5$ region.
The resulting partial phase diagram is shown in Fig. 4.

\section  {Discussions }

We have studied the behaviour of essentially three quantities and found that 
they carry information about the quantum phase transitions  in the Ising
and ANNNI models in the interface approach.
 Of these, the behaviour of the total interface 
energy had been known earlier, but the scaling of the interfacial
cooperative energy  and interfacial magnetisation appear to be
new results. However, there were earlier evidence that 
 the cooperative energy contribution is significant in a study of 
quantum spin glasses [1,9].

 In  [4], it was argued that one should look at the
scaling behaviour of the quantity $L \Delta E$ which 
is expected to have a stiffness exponent  = 1 for the transverse 
Ising chain (the same as that  of the 2-$d$ classical
model). However, this is the same as saying 
$\Delta E$ scales as $L^0$, and we do not find this behaviour (except, of 
course, at $\Gamma = 0$,  but  we are interested in the scaling 
behaviour  near the critical point). On the other hand, we do find that
  $\Delta E^c$ does have a 
stiffness exponent 0, (i.e. scales as $L^0$)   while $ \Delta E $ shows a  scaling bahaviour 
with a stiffness exponent  $= - 1$ (see Figs. 1(a-c) drawn with
$\nu = 1$).

Now, in case of the classical systems, we have  stiffness exponent = $d-1$. Of 
course for  $d = 1$, 
there is no thermal
phase transition and therefore the exponent $\alpha = 0$ is never encountered.
But, here we do have a phase transition driven by quantum fluctuations 
and that may be the reason for obtaining  an exponent $\alpha$\ = 0 for  the 
interfacial cooperative energy. 
The interfacial  magnetisation also scales with an exponent $\alpha = 0$.
The scaling function $g_c(x)$ for the interfacial cooperative 
energy is also evidently of the following form

$$g_c(x) = a  ~~\rm{for}~~ x < 0$$
$$~~~~~~~~= 0  ~~\rm{for}~~ x > 0.$$
where $a$ is a constant depending on $\kappa$.
It maybe noted that the magnetisation depends  not only on 
the number of interfaces but also their positions and it is apparent  from the data 
that as the system size is increased, the interface caused by the 
twist moves towards
the center of the chain.  Therefore, the exponent $\alpha =0$ for 
the interfacial magnetisation is not surprising.

 One  can say that the nontrivial exponent 
of $- 1$ obtained for the
total interfacial energy is  a novel feature of the quantum model. 
On the other hand, if one looks at the scaling functions  in Figs. 1-3, 
it is obvious  that they are different for $L \Delta E$ and $\Delta E^c$.
The scaling functions for $\Delta E^c$ and $\Delta m$ are, however,
similar.  Apparently the scaling function $g(x)$ for 
$L\Delta E $ has the following form 

$$g(x) \sim -x ~~\rm{for}~~ x < 0$$
$$~~~~~~~~= 0  ~~\rm{for}~~ x > 0$$
such that   $L\Delta E \sim (\Gamma_c - \Gamma )L $
which is the expected behaviour mentioned in [4].

 The  scaling behaviour of 
$\Delta E^c$ and $\Delta E$ are different but  
 the quantities $\Delta E^c$ and $L\Delta E$  have 
the same stiffness exponent. 
Hence, there is an additional dimension $L$ in the  total energy  which may 
be related to the  
   additional dimension which comes into play in  quantum models.

That the interface method is quite powerful is again proved. 
We obtain the exact critical point for the 
transverse Ising chain and  a phase diagram for the transverse ANNNI model 
consistent with
the previous studies.
However, we did not venture to investigate the regime $\kappa > 0.5$
in the ANNNI model 
 fully because of the nontrivial nature of the 
transition to a possible floating phase. 
 The phase boundary where the antiphase disappears is 
also not obtained for $\kappa < 0.7 $ because of the difficulty
in imposing conflicting boundary conditions. Since in degenerate 
systems, there can be a number of ways to impose the FBC and the UBC
we tried several combinations but faced the 
same difficulty. This is because of the
very structure of the degenerate ground states as elaborated in 
section II. It is true that the more interesting phase transitions
for $\kappa > 0.5$ could not be obtained here, but we showed  that 
estimating the boundary above which the antiphase vanishes is 
 a nontrivial task itself.  In fact,  most of the analytical and numerical 
methods give an incomplete picture for $\kappa > 0.5$. 

\vskip 2cm

\noindent {\bf Acknowledgments} 
\vskip 1cm
This work is supported by SFB341.
The author is  grateful to Subinay Dasgupta for bringing ref. [4] to notice,
and also very much to Heiko Rieger for useful discussions during the 
development of the program and  for sending   preprint of ref. [8]
prior to publication. She is thankful to  Dietrich Stauffer for discussions.

\pagebreak

\pagebreak
\begin{figure}

\caption  {Plot of (a) $\Delta E L$, (b) $\Delta E^c$ and (c) $\Delta m$ 
 vs $x = (\Gamma -\Gamma_c)L^{1/\nu}$ for system sizes 
10 ($\Diamond $ ), 12 ($+$), 14 ($\Box$), 16($\times$) and 20 ($\triangle$) for the 
Ising chain in transverse field with $\Gamma _c$ = 1 and $\nu = 1$ ($E, \Gamma $
 in units of $J$).}
\end {figure}

\begin{figure}

\caption  {Plot of (a) $\Delta E L$, (b) $\Delta E^c$ and (c) $\Delta m$ 
 vs $x = (\Gamma -\Gamma_c)L^{1/\nu}$ for system sizes 
10 ($\Diamond $ ), 12 ($+$), 14 ($\Box$), 16($\times$) and 20 
($\triangle$) for the 
ANNNI  chain in transverse field with $\Gamma _c$ = 0.42 and $\nu = 1$ at 
$\kappa $ = 0.3 ($E, \Gamma $
 in units of $J$).}
\end {figure}
\begin{figure}
\caption  {Plot of (a) $\Delta E L$ and  (b) $\Delta E^c$  
 vs $x = (\Gamma -\Gamma_c)L^{1/\nu}$ for system sizes 
8 ($\Diamond $ ), 12 ($+$), 16 ($\Box$) and  20 ($\times$) 
 for the 
ANNNI  chain in transverse field with $\Gamma _c$ = 0.52 and $\nu = 1$ at 
$\kappa $ = 1.0 ($E, \Gamma $
 in units of $J$).}
\end{figure}

\begin{figure}
\caption  {Partial phase diagram for ANNNI chain in transverse field : for
frustration parameter $\kappa < 0.5$ the data points indicate 
the critical fields for
the ferromagnetic ($F$) to paramagnetic ($P$) transition, for $ \kappa  > 0.5$, the
data points indicate the disappearance of the antiphase ($<2>$.) }
\end{figure}
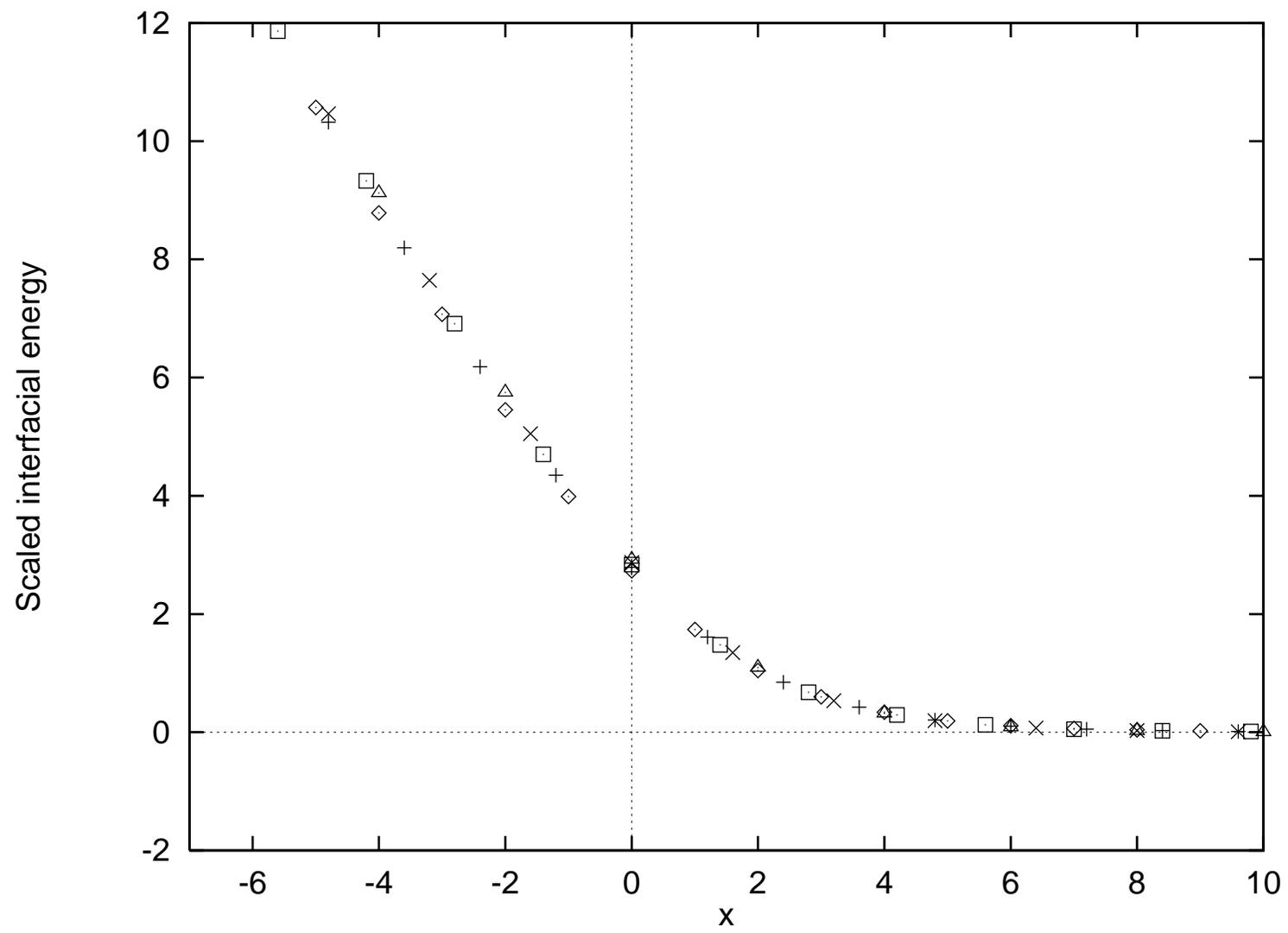
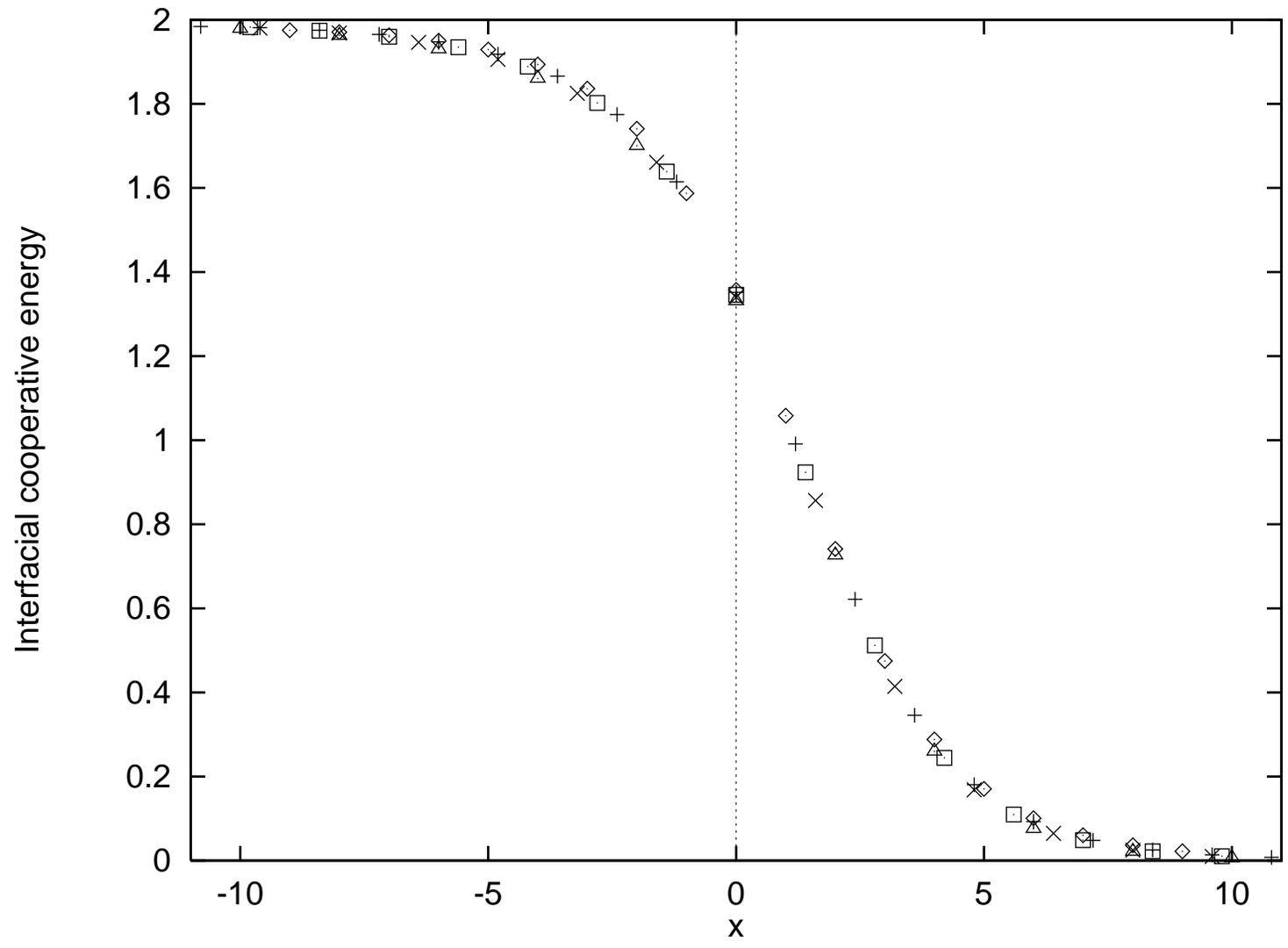
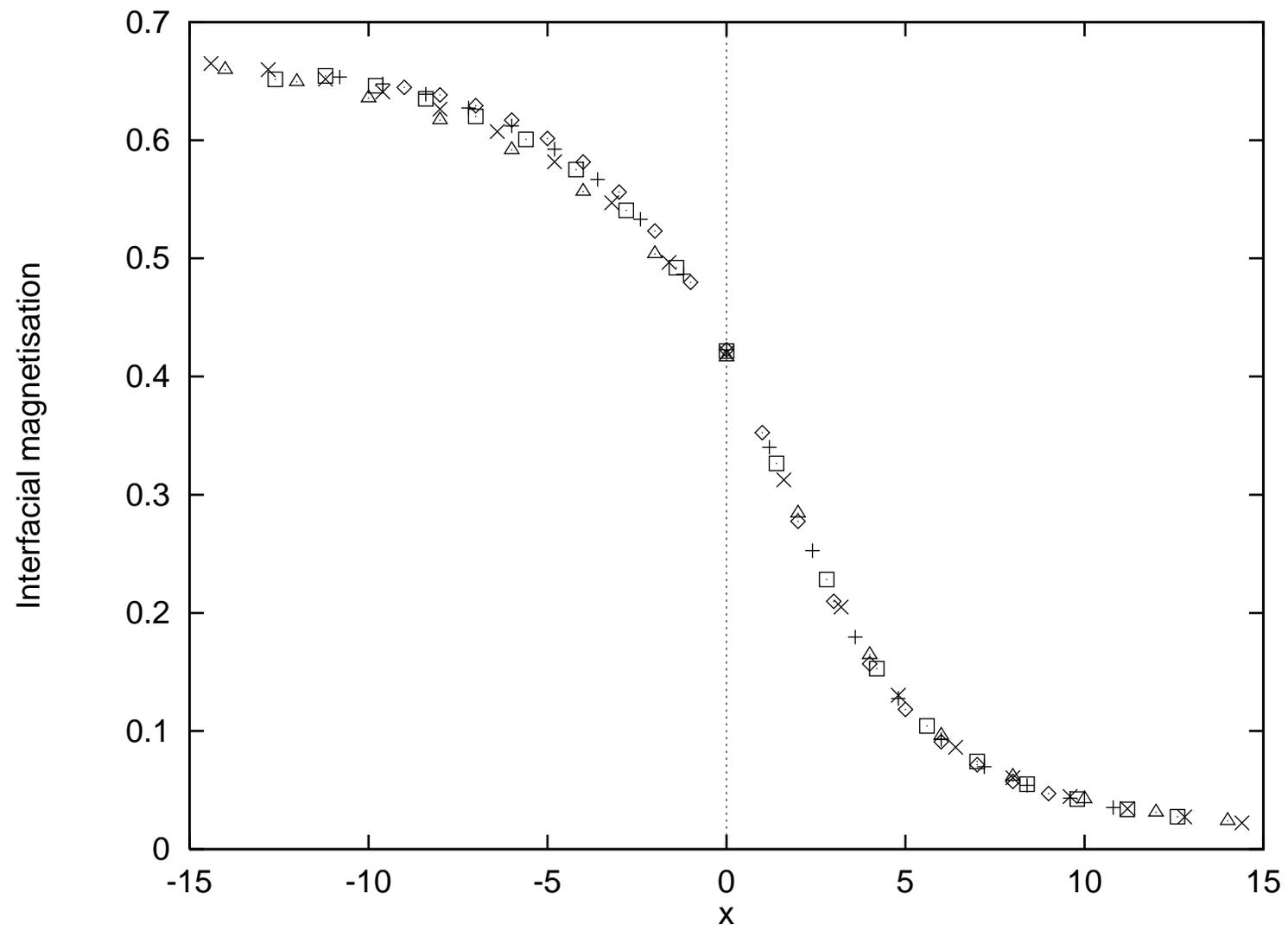
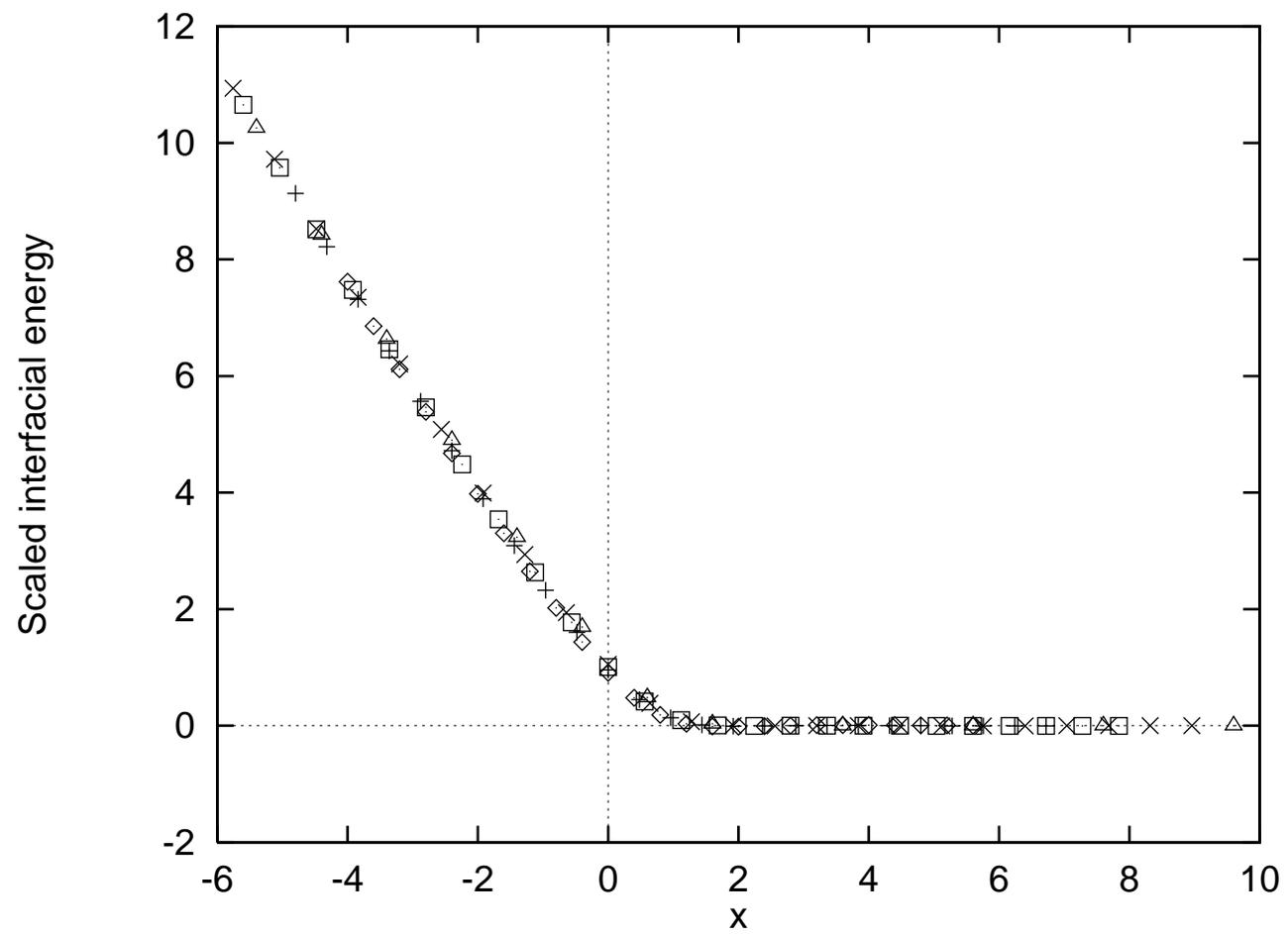
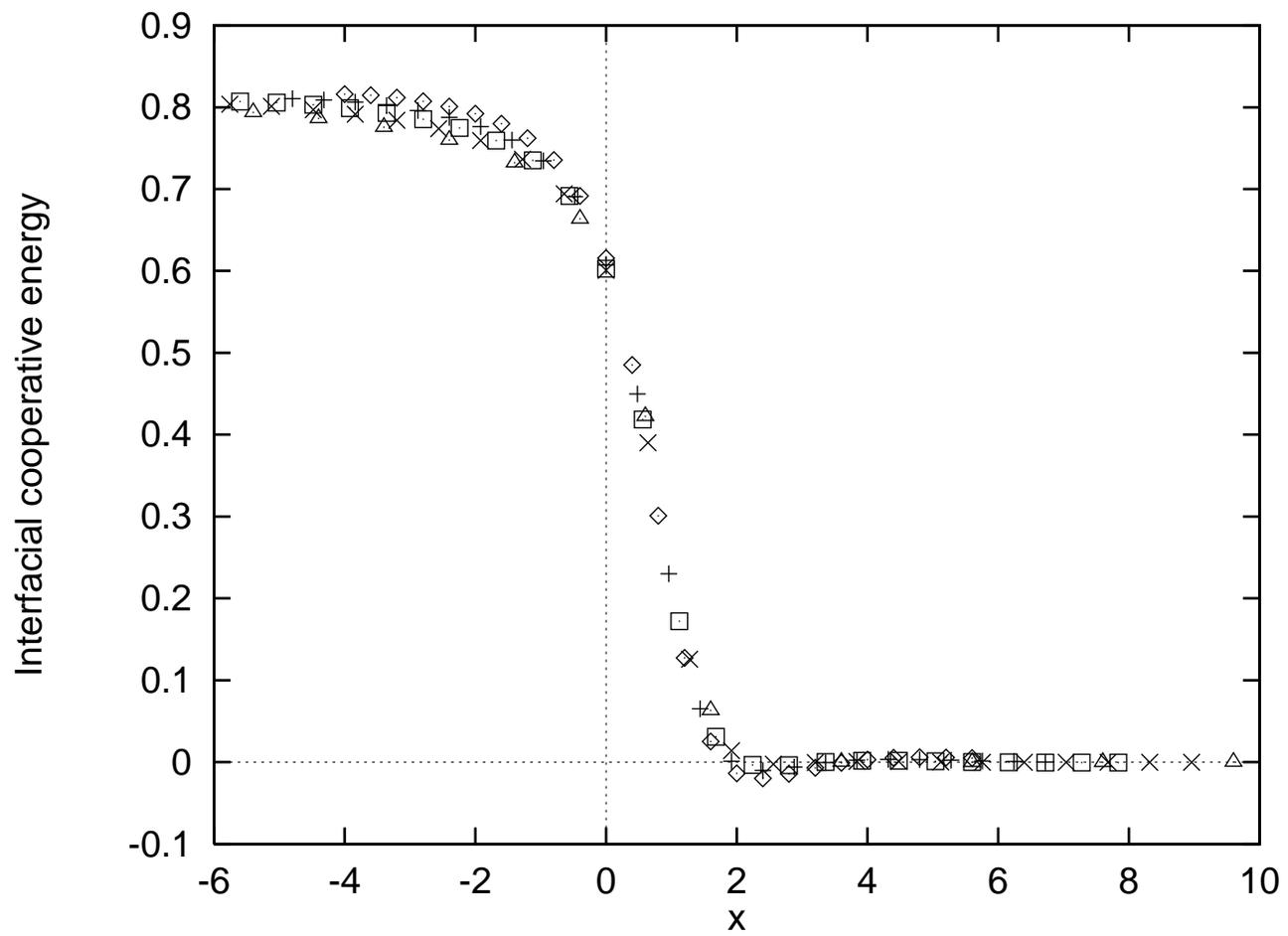
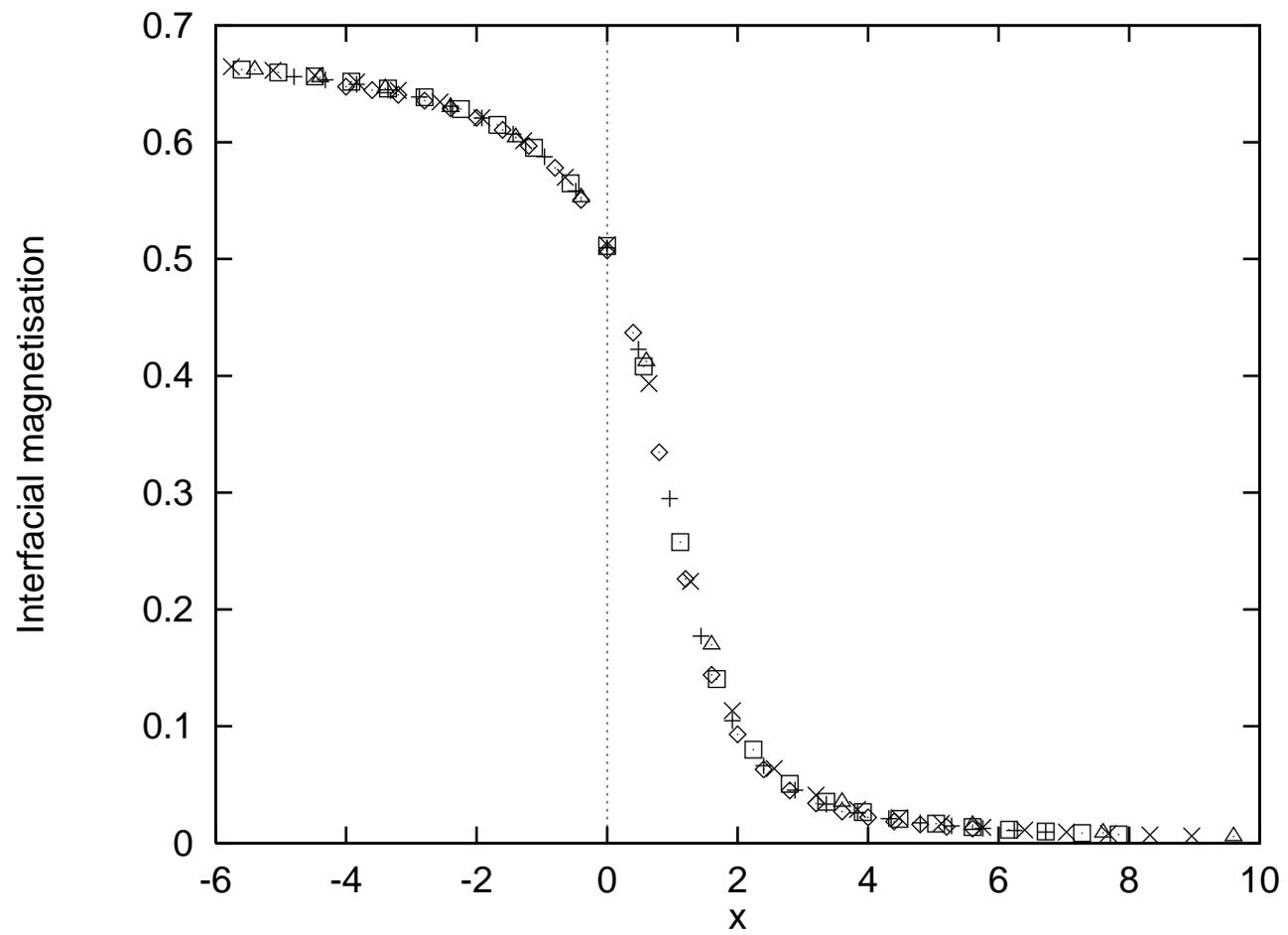
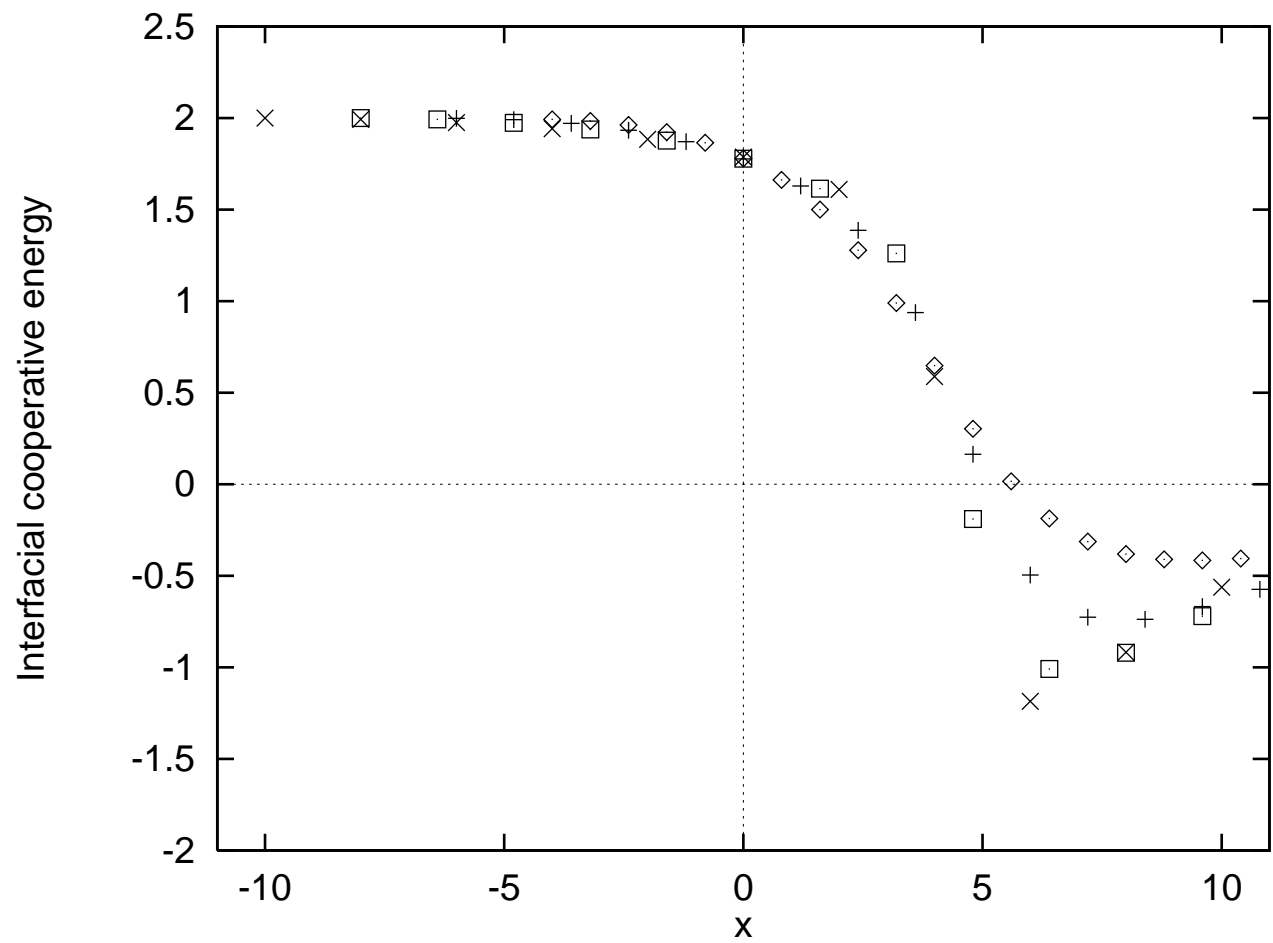
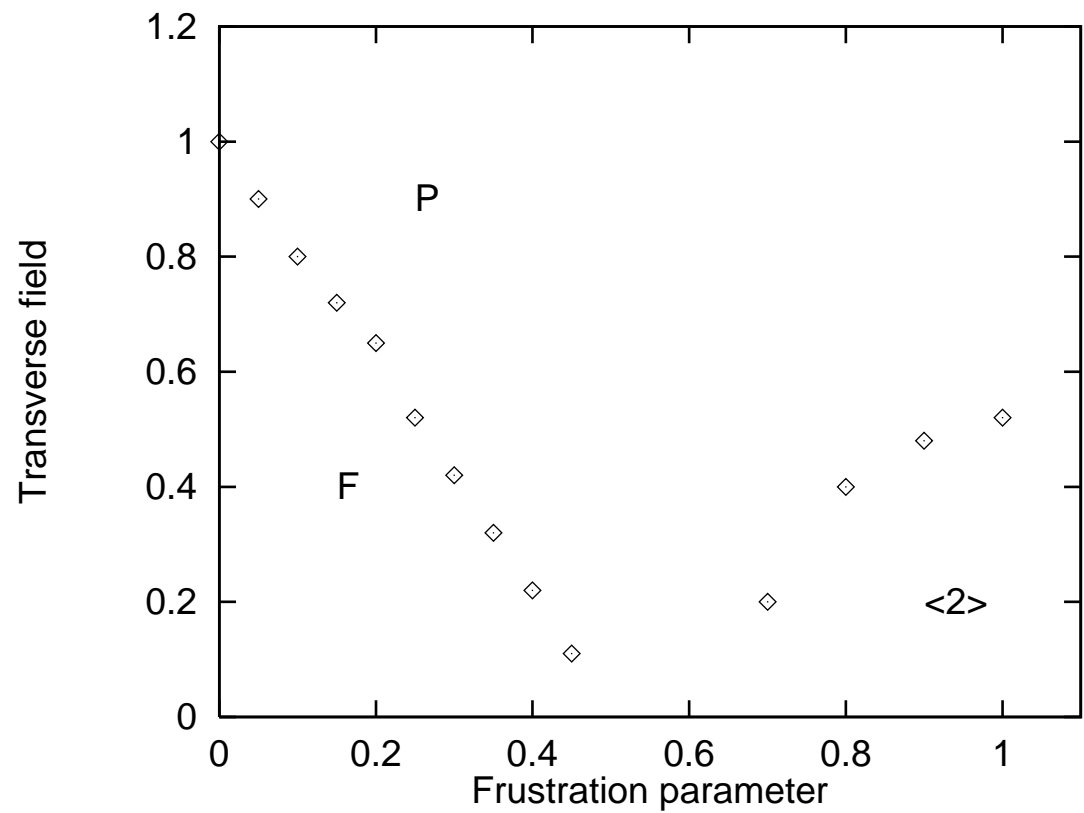
\end{document}